\documentclass[aps,twocolumn]{revtex4}
\usepackage{amsmath,amssymb}

\begin{document}

\title{Interacting agegraphic dark energy models in non-flat universe}

\author{Ahmad Sheykhi \footnote{
sheykhi@mail.uk.ac.ir}}
\address{Department of Physics, Shahid Bahonar University, P.O. Box 76175, Kerman, Iran\\
         Research Institute for Astronomy and Astrophysics of Maragha (RIAAM), Maragha,
         Iran}

\begin{abstract}
A so-called ``agegraphic dark energy'', was recently proposed to
explain the dark energy-dominated universe. In this Letter, we
generalize the agegraphic dark energy models to the universe with
spatial curvature in the presence of interaction between dark
matter and dark energy. We show that these models can accommodate
$w_D = -1 $ crossing for the equation of state of dark energy. In
the limiting case of a flat universe, i.e. $k = 0$, all previous
results of agegraphic dark energy in flat universe are restored.

\end{abstract}
\maketitle
\section{Introduction\label{Int}}
The dark energy problem constitute a major puzzle of modern
cosmology. A great variety of cosmological observations suggest
that our universe is currently undergoing a phase of accelerated
expansion likely driven by some unknown energy component whose
main feature is to possess a negative pressure \cite{Rie}.
Although the nature of such dark energy is still speculative, an
overwhelming flood of papers has appeared which attempt to
describe it by devising a great variety of models. Among them are
cosmological constant, exotic fields such as phantom or
quintessence, modified gravity, etc, see \cite{Pad} for a recent
review. Recently, a new dark energy candidate, based not in any
specific field but on the holographic principle, was proposed
\cite{Hor,Hsu}. According to the holographic principle, the number
of degrees of freedom of physical systems scale with their
bounding area rather than with their volume \cite{Suss1}. On these
basis, Cohen et al. \cite{Coh} suggested that in quantum field
theory a short distance cutoff is related to a long distance
cutoff due to the limit set by formation of a black hole, which
results in an upper bound on the zero-point energy density. The
extension of the holographic principle to a general cosmological
setting was addressed by Fischler and Susskind \cite{Suss2}.
Following this line, Li \cite{Li} argued that zero-point energy
density could be viewed as the holographic dark energy density
satisfying $\rho_D \leq3c^2m^2_p/L^2$, the equality sign holding
only when the holographic bound is saturated. Here $c^2$ is a
constant, the coefficient $3$ is for convenience, $L$ is an IR
cutoff and $m^2_p =(8\pi G)^{-1}$.  The holographic models of dark
energy have been proposed and studied widely in the literature
\cite{Huang,HDE,Xin,Setare}. It is fair to claim that simplicity
and reasonability of holographic model of dark energy provides
more reliable frame to investigate the problem of dark energy
rather than other models proposed in the literature \cite{Seta2}.
For example the coincidence problem can be easily solved in some
models of holographic dark energy \cite{Pav1}.

More recently, a new dark energy model, called agegraphic dark
energy (ADE) was proposed by Cai \cite{Cai1}. This model is based
on the uncertainty relation of quantum mechanics together with the
gravitational effect in general relativity. Following the line of
quantum fluctuations of spacetime, Karolyhazy et al. \cite{Kar1}
argued that the distance $t$ in Minkowski spacetime cannot be
known to a better accuracy than $\delta{t}=\beta
t_{p}^{2/3}t^{1/3}$ where $\beta$ is a dimensionless constant of
order unity. Based on Karolyhazy relation, Maziashvili discussed
that the energy density of metric fluctuations of the Minkowski
spacetime is given by \cite{Maz}
\begin{equation}\label{rho0}
\rho_{D} \sim \frac{1}{t_{p}^2 t^2} \sim \frac{m^2_p}{t^2},
\end{equation}
where $t_{p}$ is the reduced Planck time. Throughout this Letter
we use the units $c =\hbar=k_b = 1$. Therefore one has $l_p = t_p
= 1/m_p$ with $l_p$ and $m_p$ are the reduced Planck length and
mass, respectively. The agegraphic dark energy model assumes that
the observed dark energy comes from the spacetime and matter field
fluctuations in the universe \cite{Wei1,Wei2}. The agegraphic
models of dark energy  have been examined \cite{age1,shey1,seta3}
and constrained by various astronomical observations \cite{age2}.

On the other hand, lacking a fundamental theory, most discussions
on dark energy rely on the fact that its evolution is independent
of other matter fields. Given the unknown nature of both dark
matter and dark energy there is nothing in principle against their
mutual interaction and it seems very special that these two major
components in the universe are entirely independent. Indeed, this
possibility is receiving growing attention in the literature
\cite{Ame,Zim} (see also \cite{Seta1,wang1} and references
therein) and appears to be compatible with SNIa and CMB data
\cite{Oli}. Furthermore, the interacting holographic dark energy
models have also been extended to the universe with spacial
curvature \cite{Seta2,wang2}.

Besides, it is generally believed that inflation practically
washes out the effect of curvature in the early stages of cosmic
evolution. However, it does not necessarily imply that the
curvature has to be wholly neglected at present. Indeed, aside
from the sake of generality, there are sound reasons to include
it: (i) Inflation drives the $k/a^2$ ratio close to zero but it
cannot set it to zero if $k\neq 0$ initially. (ii) The closeness
to perfect flatness depends on the number of e-folds and we can
only speculate about the latter. (iii) After inflation the
absolute value of the $k/a^2$ term in the field equations may
increase with respect to the matter density term, thereby the
former should not be ignored when studying the late universe. (iv)
Observationally there is room for a small but non-negligible
spatial curvature \cite{spe}. For instance, the tendency of
preferring a closed universe appeared in a suite of CMB
experiments \cite{Sie}. The improved precision from WMAP provides
further confidence, showing that a closed universe with positively
curved space is marginally preferred \cite{Uzan}. In addition to
CMB, recently the spatial geometry of the universe was probed by
supernova measurements of the cubic correction to the luminosity
distance \cite{Caldwell}, where a closed universe is also
marginally favored.

In the light of all mentioned above, it becomes obvious that the
investigation on the interacting agegraphic dark energy models in
the universe with spacial curvature is well motivated. In this
Letter, we would like to generalize, following \cite{Seta2}, the
agegraphic dark energy models to the universe with spacial
curvature in the presence of interaction between the dark matter
and dark energy. We will also show that the equation of state of
dark energy can accommodate $w = -1 $ crossing. The plan of the
work is as follows: In section \ref{ORI}, we study the original
agegraphic model of dark energy in a non-flat universe where the
time scale is chosen to be the age of the universe. In section
\ref{NEW}, we consider the new model of agegraphic dark energy
while the time scale is chosen to be the conformal time instead of
the age of the universe. Finally, in section \ref{CONC} we
summarize our results.

\section{THE ORIGINAL ADE IN NONFLAT UNIVERSE\label{ORI}}
The original agegraphic dark energy density has the form
(\ref{rho0}) where $t$ is chosen to be the age of the universe
\begin{equation}
T=\int{dt}=\int_0^a{\frac{da}{Ha}},
\end{equation}
where $a$ is the scale factor and $H=\dot{a}/a$ is the Hubble
parameter. Thus, the energy density of the agegraphic dark energy
is given by \cite{Cai1}
\begin{equation}\label{rho1} \rho_{D}=
\frac{3n^2 m_{p}^2}{T^2},
\end{equation}
where the numerical factor $3n^2$ is introduced to parameterize
some uncertainties, such as the species of quantum fields in the
universe, the effect of curved space-time (since the energy
density is derived for Minkowski space-time), and so on. The dark
energy density (\ref{rho1}) has the same form as the holographic
dark energy, but  the length measure is chosen to be the age of
the universe instead of the horizon radius of the universe. Thus
the causality problem in the holographic dark energy is avoided
\cite{Cai1}.

The total energy density is $\rho=\rho_{m}+\rho_{D}$, where
$\rho_{m}$ and $\rho_{D}$ are the energy density of dark matter
and dark energy, respectively. The total energy density satisfies
a conservation law
\begin{equation}\label{cons}
\dot{\rho}+3H(\rho+p)=0.
\end{equation}
However, since we consider the interaction between dark matter and
dark energy, $\rho_{m}$ and $\rho_{D}$ do not conserve separately;
they must rather enter the energy balances
\begin{eqnarray}
&&\dot{\rho}_m+3H\rho_m=Q, \label{consm}
\\&& \dot{\rho}_D+3H\rho_D(1+w_D)=-Q.\label{consq}
\end{eqnarray}
Here $w_D$ is the equation of state parameter of agegraphic dark
energy and $Q$ denotes the interaction term and can be taken as $Q
=3b^2 H\rho$  with $b^2$  being a coupling constant. This
expression for the interaction term was first introduced in the
study of the suitable coupling between a quintessence scalar field
and a pressureless cold dark matter field \cite{Ame}. In the
context of holographic dark energy model, this form of interaction
was derived from the choice of Hubble scale as the IR cutoff
\cite{Pav1}. Although at this point the interaction may look
purely phenomenological but different Lagrangians have been
proposed in support of it (see \cite{Tsu} and references therein).
Besides, in the absence of a symmetry that forbids the interaction
there is nothing, in principle, against it. Further, the
interacting dark mater–dark energy (the latter in the form of a
quintessence scalar field and the former as fermions whose mass
depends on the scalar field) has been investigated at one quantum
loop with the result that the coupling leaves the dark energy
potential stable if the former is of exponential type but it
renders it unstable otherwise \cite{Dor}. Thus, microphysics seems
to allow enough room for the coupling; however, this point is not
fully settled and should be further investigated. The difficulty
lies, among other things, in that the very nature of both dark
energy and dark matter remains unknown whence the detailed form of
the coupling cannot be elucidated at this stage.

For the Friedmann-Robertson-Walker (FRW) universe filled with dark
energy and dust (dark matter), the corresponding Friedmann
equation takes the form
\begin{eqnarray}\label{Fried}
H^2+\frac{k}{a^2}=\frac{1}{3m_p^2} \left( \rho_m+\rho_D \right),
\end{eqnarray}
where $k$ is the curvature parameter with $k = -1, 0, 1$
corresponding to open, flat, and closed universes, respectively. A
closed universe with a small positive curvature
($\Omega_k\simeq0.02$) is compatible with observations \cite{spe}.
If we introduce, as usual, the fractional energy densities such as
\begin{eqnarray}\label{Omega}
\Omega_m=\frac{\rho_m}{3m_p^2H^2}, \hspace{0.5cm}
\Omega_D=\frac{\rho_D}{3m_p^2H^2},\hspace{0.5cm}
\Omega_k=\frac{k}{H^2 a^2},
\end{eqnarray}
then the Friedmann equation can be written
\begin{eqnarray}\label{Fried2}
\Omega_m+\Omega_D=1+\Omega_k.
\end{eqnarray}
Using Eq. (\ref{rho1}), we have
\begin{eqnarray}\label{Omegaq}
\Omega_D=\frac{n^2}{H^2T^2}.
\end{eqnarray}
Differentiating Eq. (\ref{Omegaq}) and using relation
${\dot{\Omega}_D}= {\Omega'_D} H$, we reach
\begin{eqnarray}\label{Omegaq2}
{\Omega'_D}=\Omega_D\left(-2\frac{\dot{H}}{H^2}-\frac{2}{n
}\sqrt{\Omega_D}\right),
\end{eqnarray}
where the dot is the derivative with respect to the cosmic time
and the prime denotes the derivative with respect to $x=\ln{a}$.
Taking the derivative of  both side of the Friedman equation
(\ref{Fried}) with respect to the cosmic time, and using Eqs.
(\ref{rho1}), (\ref{consm}), (\ref{Fried2})  and (\ref{Omegaq}),
it is easy to find that
\begin{eqnarray}\label{Hdot}
\frac{\dot{H}}{H^2}=-\frac{3}{2}(1-\Omega_D)-\frac{\Omega^{3/2}_D}{n}-\frac{\Omega_k}{2}
+\frac{3}{2}b^2(1+\Omega_k).
\end{eqnarray}
Substituting this relation into Eq. (\ref{Omegaq2}), we obtain the
equation of motion of agegraphic dark energy
\begin{eqnarray}\label{Omegaq3}
{\Omega'_D}&=&\Omega_D\left[(1-\Omega_D)\left(3-\frac{2}{n}\sqrt{\Omega_D}\right)
\right. \nonumber
\\
&& \left.-3b^2(1+\Omega_k)+\Omega_k\right].
\end{eqnarray}
Inserting $\Omega_k=0=b$, this equation reduces to Eq. (12) of
Ref. \cite{Cai1}. Using Eqs. (\ref{rho1}) and (\ref{consq}), as
well as Eq. (\ref{Omegaq}), we can obtain the equation of state
parameter for the interacting agegraphic dark energy
\begin{eqnarray}\label{wq}
w_D=-1+\frac{2}{3n}\sqrt{\Omega_D}-b^2 \Omega^{-1}_D(1+\Omega_k).
\end{eqnarray}
The total equation of state parameter is given by
\begin{eqnarray}\label{wtot}
w_{\mathrm{tot}}=\frac{p}{\rho}=\frac{\Omega_D}{1+\Omega_k}w_D.
\end{eqnarray}
For completeness, we give the deceleration parameter
\begin{eqnarray}
q=-\frac{\ddot{a}}{aH^2}=-1-\frac{\dot{H}}{H^2},\label{q0}
\end{eqnarray}
which combined with the Hubble parameter and the dimensionless
density parameters form a set of useful parameters for the
description of the astrophysical observations. Substituting Eq.
(\ref{Hdot}) in Eq. (\ref{q0}) we get
\begin{eqnarray}\label{q}
q&=&\frac{1}{2}-\frac{3}{2}{\Omega_D}
+\frac{\Omega^{3/2}_D}{n}-\frac{3}{2}b^2
+\frac{1}{2}\Omega_k(1-3b^2).
\end{eqnarray}
It is worth noting that in the absence of interaction between dark
energy and dark matter, $b^2=0$,  from Eq. (\ref{wq}) we see that
$w_D$ is always larger than $-1$ and cannot cross the phantom
divide $w_D =-1$.  In addition the condition $n>1$ is necessary to
derive the present accelerated expansion \cite{Cai1}. However, the
situation is changed as soon as the interaction term is taken into
account. In this case ($b^2\neq0$), from Eq. (\ref{wq}) one can
easily see that $w_D$ can cross the phantom divide provided
$3nb^2(1+\Omega_k)>2{\Omega^{3/2}_D}$. If we take $\Omega_D=0.72$
and  $\Omega_k=0.02$  for the present time, the phantom-like
equation of state can be achieved only if $nb^2>0.4$. The best fit
result for agegraphic dark energy  which is consistent with most
observations like WMAP and SDSS data, shows that $n = 3.4$
\cite{age2}. Thus, the condition $w_D<-1$ leads to $b^2>0.12$ for
the coupling between dark energy and dark matter. For instance, if
we take $b^2=0.15$ we get $w_D=-1.05$. This indicates that one can
generate phantom-like equation of state from an interacting
agegraphic dark energy model in the universe with any spacial
curvature. Putting $\Omega_k=0$ in Eqs. (\ref{wq}) and (\ref{q}),
these equations reduce to their respective equations of original
interacting agegraphic dark energy model in flat universe
\cite{Wei1}. The original interacting agegraphic dark energy has
laso some interesting features. From Eq. (\ref{wq}), it is easy to
see that $w_D<-1$ is necessary in the early time where
$\Omega_D\rightarrow0$, while in the late time where
$\Omega_D\rightarrow1$ ($\Omega_k\simeq0$), we have $w_D>-1$ for
$nb^2<0.67$ and $w_D<-1$ for $nb^2>0.67$.

It is important to note that in the absence of interaction, the
original agegraphic dark energy model has a drawback to describe
the matter-dominated universe in the far past where $a \ll1$ and
$\Omega_D \ll 1$. On one side, from Eq. (\ref{wq}) with $b^2=0$ we
have $w_D \rightarrow-1$ as $\Omega_D\rightarrow0$. This means
that in the matter-dominated epoch the agegraphic dark energy
behaves like a cosmological constant. On the other side, Eq.
(\ref{Omegaq3}) with $b^2=0$, $\Omega_D \ll 1$ and $\Omega_k \ll
1$ approximately becomes
\begin{eqnarray}\label{Omegaqq3}
\frac{d\Omega_D}{da}\simeq\frac{\Omega_D}{a}
\left(3-\frac{2}{n}\sqrt{\Omega_D}\right),
\end{eqnarray}
which has a solution of the form $\Omega_D =9n^2/4$. Substituting
this relation into Eq. (\ref{wq}) with $b^2=0$, we get $w_D =0$.
Therefore, the dark energy behaves as pressureless matter.
Obviously, pressureless matter cannot generate accelerated
expansion, which seems to rule out the choice $t=T$. Thus there is
a  confusion in the original agegraphic dark energy model. This
issue is similar to the holographic dark energy model when
choosing the Hubble parameter as the IR cutoff \cite{Hsu}. It was
argued by Pavon and Zimdahl that the problem can be solved as soon
as an interaction between the dark energy and dark matter is taken
into account \cite{Pav1}. Similarly, in the agegraphic model of
dark energy the inconsistency in the original version can be
removed with the interaction between dark energy and dark matter.
To see this, consider the matter-dominated epoch where $a\ll1$ and
$\Omega_D \ll1$ for interacting agegraphic dark energy. In this
case Eq. (\ref{Omegaq3}) with $\Omega_k \ll 1$ approximately
becomes
\begin{eqnarray}\label{Omegaq32}
\frac{d\Omega_D}{da}\simeq \frac{\Omega_D}{a}
\left(3-\frac{2}{n}\sqrt{\Omega_D}-3b^2\right),
\end{eqnarray}
which has a solution of the form $\Omega_D =9n^2 (1-b^2)^2/4$.
Substituting this relation into Eq. (\ref{wq}), we obtain
\begin{eqnarray}\label{wq2}
w_D=-b^2 \left(1+\frac{4}{9n^2(1-b^2)^2}\right).
\end{eqnarray}
Therefore for $b\neq0$ we have $w_D<0$,  and the agegraphic model
of dark energy can generate accelerated expansion. The presence of
the spatial curvature does not seriously modify the above
discussion. Thus, the confusion in the original agegraphic dark
energy model without interaction is removed.  Nevertheless, Wei
and Cai \cite{Wei2} proposed a new model of agegraphic dark energy
to resolve the contradiction in the far past of original
non-interacting agegraphic dark energy model.

\section{THE NEW MODEL OF ADE IN NONFLAT UNIVERSE\label{NEW}}
As we argued  the original agegraphic dark energy model has some
difficulties \cite{Cai1}. Therefore one may seek for another
agegraphic dark energy model. Wei and Cai proposed a new model of
agegraphic dark energy \cite{Wei2}, while the time scale is chosen
to be the conformal time $\eta$ instead of the age of the
universe, which is defined by $dt= ad\eta$, where $t$ is the
cosmic time. It is worth noting that the Karolyhazy relation
$\delta{t}= \beta t_{p}^{2/3}t^{1/3}$ was derived for Minkowski
spacetime $ds^2 = dt^2-d\mathrm{x^2}$ \cite{Kar1,Maz}. In the case
of the FRW universe, we have $ds^2 = dt^2-a^2d\mathrm{x^2} =
a^2(d\eta^2-d\mathrm{x^2})$. Therefore, it is more reasonable to
choose the time scale in Eq. (\ref{rho1}) to be the conformal time
$\eta$ \cite{Wei2}. Taking this into account, the energy density
of the new agegraphic dark energy can be written
\begin{equation}\label{rho3}
\rho_{D}= \frac{3n^2 m_{p}^2}{\eta^2},
\end{equation}
where the conformal time is given by
\begin{equation}
\eta=\int{\frac{dt}{a}}=\int_0^a{\frac{da}{Ha^2}}.
\end{equation}
If we write $\eta$ to be a definite integral, there will be an
integral constant in addition. Thus, we have $\dot{\eta}=1/a$. Let
us again consider a FRW universe with spatial curvature containing
the new agegraphic dark energy and pressureless matter. The
Friedmann equation can be written
\begin{eqnarray}\label{Friednew}
H^2+\frac{k}{a^2}=\frac{1}{3m_p^2} \left( \rho_m+\rho_D \right),
\end{eqnarray}
where can be rewritten as
\begin{eqnarray}\label{Fried2new}
\Omega_m+\Omega_D=1+\Omega_k.
\end{eqnarray}
The fractional energy density of the agegraphic dark energy is now
given by
\begin{eqnarray}\label{Omegaqnew}
\Omega_D=\frac{n^2}{H^2\eta^2}.
\end{eqnarray}
We can also find the equation of motion for $\Omega_D$ by taking
the derivative of Eq. (\ref{Omegaqnew}). The result is
\begin{eqnarray}\label{Omegaq2new}
{\Omega'_D}=\Omega_D\left(-2\frac{\dot{H}}{H^2}-\frac{2}{na
}\sqrt{\Omega_D}\right).
\end{eqnarray}
Taking the derivative of both side of the Friedman equation
(\ref{Friednew}) with respect to the cosmic time $t$, and using
Eqs. (\ref{consm}), (\ref{rho3}), (\ref{Fried2new})  and
(\ref{Omegaqnew}), we obtain
\begin{eqnarray}\label{Hdotnew}
\frac{\dot{H}}{H^2}=-\frac{3}{2}(1-\Omega_D)-\frac{\Omega^{3/2}_D}{na}-\frac{\Omega_k}{2}
+\frac{3}{2}b^2(1+\Omega_k).
\end{eqnarray}
Substituting this relation into Eq. (\ref{Omegaq2new}), we obtain
the evolution behavior of the new agegraphic dark energy
\begin{eqnarray}\label{Omegaq3new}
{\Omega'_D}&=&\Omega_D\left[(1-\Omega_D)\left(3-\frac{2}{na}\sqrt{\Omega_D}\right)
\right. \nonumber\
\\
&& \left.-3b^2(1+\Omega_k)+\Omega_k\right].
\end{eqnarray}
The equation of state parameter of the  interacting  new
agegraphic dark energy can be obtained as
\begin{eqnarray}\label{wqnew}
w_D=-1+\frac{2}{3na}\sqrt{\Omega_D}-b^2 \Omega^{-1}_D(1+\Omega_k),
\end{eqnarray}
Again we see that $w_D$ can cross the phantom divide provided
$3nab^2(1+\Omega_k)>2{\Omega^{3/2}_D}$. Taking $\Omega_D=0.72$,
$\Omega_k=0.02$, and $a=1$  for the present time, the phantom-like
equation of state can be achieved only if $nb^2>0.4$. The joint
analysis of the astronomical data for the new agegraphic dark
energy gives the best-fit value (with $1\sigma$ uncertainty) $n =
2.7$ \cite{age2}. Thus, the condition $w_D<-1$ leads to $b^2>0.15$
for the coupling between dark energy and dark matter. For
instance, if we take $b^2=0.2$ we get $w_D=-1.07$. The
deceleration parameter is now given by
\begin{eqnarray}\label{qnew}
q&=&-1-\frac{\dot{H}}{H^2}=\frac{1}{2}-\frac{3}{2}{\Omega_D}
+\frac{\Omega^{3/2}_D}{na}\nonumber \\&&-\frac{3}{2}b^2
+\frac{1}{2}\Omega_k(1-3b^2).
\end{eqnarray}
Comparing Eqs. (\ref{Hdotnew})-(\ref{qnew}) with Eqs.
(\ref{Hdot}), (\ref{Omegaq3}), (\ref{wq}) and (\ref{q}) in the
previous section, we see that the scale factor $a$ enters Eqs.
(\ref{Hdotnew})-(\ref{qnew}) explicitly. In the late time where
$a\rightarrow\infty$ and $\Omega_D \rightarrow1$, from Eq.
(\ref{wqnew}) with $b^2=0$  we have $w_D \rightarrow-1$; thus the
new agegraphic dark energy mimics a cosmological constant in the
late time. Let us now consider the matter-dominated epoch where
$a\ll1$ and $\Omega_D \ll1$. In this case Eq. (\ref{Omegaq3new})
with $b^2=0$ and $\Omega_k \ll 1$ approximately becomes
\begin{eqnarray}\label{Omegaq3new2}
\frac{d\Omega_D}{da}\simeq\frac{\Omega_D}{a}
\left(3-\frac{2}{na}\sqrt{\Omega_D}\right).
\end{eqnarray}
Solving this equation we find $\Omega_D =n^2 a^2/4$. Substituting
this relation into Eq. (\ref{wqnew}) with $b^2=0$, we obtain
$w_D=-2/3$. On the other hand, in the matter-dominated epoch,
$H^2\propto\rho_m\propto a^{-3}$. So $\sqrt{a}da\propto
dt=ad\eta$. Thus $\eta\propto \sqrt{a}$. From Eq. (\ref{rho1}) we
have $\rho_D\propto a^{-1}$. Putting this in conservation law,
$\dot{\rho}_D+3H\rho_D(1+w_D)=0$, we obtain $w_D=-{2}/{3}$.
Substituting this $w_D$ in Eq. (\ref{wqnew}) with $b^2=0$ and
$\Omega_k \ll 1$ we find that $\Omega_D=n^2a^2/4$ as expected.
Therefore, all things are consistent. The confusion in the
original agegraphic dark energy model does not exist in this new
model. These results are regardless of the value of $n$. Again one
can see that in the absence of interaction between dark energy and
dark matter, $b^2=0$, $w_D$ in Eq. (\ref{wqnew}) is always larger
than $-1$ and cannot cross the phantom divide. However, in the
presence of interaction, $b^2\neq0$, it is quite possible that
$w_D$ cross the phantom divide. In the limiting case $\Omega_k=0$,
Eqs. (\ref{Omegaq3new})-(\ref{qnew}), restore their respective
equations in interacting new agegraphic dark energy model in flat
universe \cite{Wei2}.
\section{Conclusions\label{CONC}}
There is a wide consensus among cosmologists that the universe has
entered a phase of accelerated expansion likely driven by dark
energy. However, the nature and the origin of such dark energy is
still the source of much debate. Indeed, until now we don't know
what might be the best candidate for dark energy to explain the
accelerated expansion. Therefore, cosmologists have attended to
various models of dark energy by considering all the possibilities
they have. In this regard, based on the uncertainty relation of
quantum mechanics together with the gravitational effect in
general relativity, Cai proposed an agegraphic dark energy model
to explain the acceleration of the cosmic expansion \cite{Cai1}.
However, the original agegraphic dark energy model had some
difficulties. In particular it fails to describe the
matter-dominated epoch properly \cite{Cai1}. Thus, Wei and Cai
\cite{Wei2} proposed a new model of agegraphic dark energy, while
the time scale is chosen to be the conformal time $\eta$ instead
of the age of the universe. In this Letter, we extended these
agegraphic dark energy models, in the presence of interaction
between dark energy and dark matter, to the universe with spatial
curvature. Although it is believed that our universe is spatially
flat, a contribution to the Friedmann equation from spatial
curvature is still possible if the number of e-foldings is not
very large \cite{Huang}. Besides, some experimental data has
implied that our universe is not a perfectly flat universe and
recent papers have favored the universe with spatial curvature
\cite{spe}. We obtained the equation of state for interacting
agegraphic energy density in a non-flat universe. When the
interaction between dark matter and agegraphic dark energy is
taken into account, the equation of state parameter of dark
energy, $w_D$, can cross the phantom divide in the universe with
any spacial curvature.

In interacting agegraphic models of dark energy, the properties of
agegraphic dark energy is determined by the parameters $n$ and $b$
together. These parameters would be obtained by confronting with
cosmic observational data. In this work we just restricted
ourselves to limited observational data. Giving the wide range of
cosmological data available, in the future we expect to further
constrain our model parameter space and test the viability of our
model.

Finally I would like to mention that, as we found, the spatial
curvature does not seriously modify the qualitative picture of the
interacting agegraphic dark energy models but it may however
affect the time of onset of the acceleration. For an open universe
($\Omega_k <0$) the acceleration sets in earlier whereas in a
closed universe ($\Omega_k >0$) the accelerated phase is delayed.
\acknowledgments{This work has been supported by Research
Institute for Astronomy and Astrophysics of Maragha, Iran.}


\begin{thebibliography}{99}
\bibitem{Rie} A.G. Riess, et al., Astron. J.  116 (1998)
1009;\\
  S. Perlmutter, et al.,  Astrophys. J.  517 (1999) 565;\\
  S. Perlmutter, et al.,  Astrophys. J.  598 (2003) 102;\\
  P. de Bernardis, et al.,  Nature  404 (2000) 955.

\bibitem{Pad} T. Padmanabhan, Phys. Rep.  380 (2003) 235;\\
P. J. E. Peebles,  B. Ratra,  Rev. Mod. Phys. 75 (2003) 559;\\
E. J. Copeland,  M. Sami and S. Tsujikawa, hep-th/060305.

\bibitem{Hor} P. Horava, D. Minic, Phys. Rev. Lett. 85 (2000) 1610; \\ P.
Horava, D. Minic, Phys. Rev. Lett. 509 (2001) 138;\\ K. Enqvist,
M. S. Sloth, Phys. Rev. Lett. 93 (2004) 221302.


\bibitem{Hsu} S. D. H. Hsu, Phys. Lett. B 594 (2004) 13.

\bibitem{Suss1}  G. 't Hooft, gr-qc/9310026;\\ L. Susskind, J. Math. Phys. 36 (1995)
6377.

\bibitem{Coh}  A. Cohen, D. Kaplan, A. Nelson, Phys. Rev. Lett. 82 (1999)
4971.


\bibitem{Suss2} W. Fischler, L. Susskind,hep-th/9806039.


\bibitem{Li} M. Li, Phys. Lett. B 603 (2004) 1.
\bibitem{Huang} Q. G. Huang, M. Li, JCAP 0408 (2004) 013.

\bibitem{HDE} E. Elizalde, S. Nojiri, S.D.
Odintsov, P. Wang, Phys. Rev. D 71 (2005) 103504;\\  B. Guberina,
R. Horvat, H. Stefancic, JCAP 0505 (2005) 001;\\ B. Guberina, R.
Horvat, H. Nikolic, Phys. Lett. B 636 (2006) 80;\\ H. Li, Z. K.
Guo, Y. Z. Zhang, Int. J. Mod. Phys. D 15 (2006) 869;
\\ Q. G. Huang, Y. Gong, JCAP 0408 (2004) 006;\\
J. P. B. Almeida, J. G. Pereira, Phys. Lett. B 636 (2006) 75;
\\  Y. Gong, Phys. Rev. D 70 (2004) 064029; \\ B. Wang, E.
Abdalla, R. K. Su, Phys. Lett. B 611 (2005) 21;\\ J. Y. Shen, B.
Wang, E. Abdalla, R. K. Su, Phys. Lett. B 609 (2005) 200.

\bibitem{Xin} X. Zhang, F. Q. Wu,  Phys. Rev. D 72 (2005)
043524;\\ X. Zhang, Phys. Rev. D 74 (2006) 103505;\\ X. Zhang, F.
Q.  Wu, Phys. Rev. D 76 (2007) 023502.

\bibitem{Setare} M. R. Setare, S. Shafei, JCAP 09 (2006) 011;\\ M. R.  Setare, Phys. Lett. B 644 (2007) 99;\\  M. R. Setare,
E. C. Vagenas, Phys. Lett. B 666 (2008) 111;\\ M. R. Setare, Phys.
Lett. B 642  (2006) 421;\\ H. M. Sadjadi, arXiv:0902.2462.



\bibitem{Seta2} M. R. Setare, Phys. Lett. B 642 (2006) 1.


\bibitem{Pav1} D. Pavon, W. Zimdahl, Phys. Lett. B 628 (2005)
206;\\
N. Banerjee, D. Pavon, Phys. Lett. B 647 (2007) 477.

\bibitem{Cai1} R. G. Cai, Phys. Lett. B 657 (2007) 228.

\bibitem{Kar1} F. Karolyhazy,
Nuovo.Cim. A 42, 390 (1966);\\ F. Karolyhazy, A. Frenkel and B.
Lukacs, in \textit{Physics as natural Philosophy} edited by A.
Shimony and H. Feschbach, MIT Press, Cambridge, MA, (1982);\\ F.
Karolyhazy, A. Frenkel and B. Lukacs, in \textit{Quantum Concepts
in Space and Time} edited by R. Penrose and C.J. Isham, Clarendon
Press, Oxford, (1986).

\bibitem{Maz} M. Maziashvili Int. J. Mod. Phys. D 16 (2007) 1531;\\ M. Maziashvili, Phys. Lett. B 652 (2007) 165.


\bibitem{Wei1} H. Wei and R. G. Cai, Eur. Phys. J. C 59 (2009) 99.

\bibitem{Wei2} H. Wei and R. G. Cai, Phys. Lett. B 660 (2008) 113.


\bibitem{age1} J. Cui, et al., arXiv:0902.0716; \\ Y. W. Kim, et al.,  Mod. Phys. Lett. A 23 (2008) 3049;\\
Y. Zhang, et al. arXiv:0708.1214;\\ J .P Wu, D. Z. Ma, Y. Ling,
Phys.
Lett. B 663,  (2008) 152; \\ K. Y. Kim, H. W. Lee, Y. S. Myung, Phys.Lett. B 660 (2008) 118;\\
\\ J. Zhang, X. Zhang, H. Liu, Eur.
Phys. J. C 54 (2008) 303; \\ I. P. Neupane, Phys. Lett. B 673
(2009) 111.

\bibitem{shey1} A. Sheykhi, arXiv:0907.5144;\\ A. Sheykhi,
0908.0606.

\bibitem{seta3} M. R. Setare, arXiv:0907.4910;\\ M. R. Setare,
arXiv:0908.0196.

\bibitem{age2} H. Wei and R. G. Cai, Phys. Lett. B 663 (2008) 1;\\
X. Wu, et al., arXiv:0708.0349.



\bibitem{Ame} L. Amendola, Phys. Rev. D 60 (1999)  043501; \\ L. Amendola, Phys. Rev. D 62 (2000) 043511;
 \\ L. Amendola and C. Quercellini, Phys. Rev. D 68
(2003)  023514; \\ L. Amendola and D. Tocchini-Valentini, Phys.
Rev. D 64 (2001)  043509 ;\\ L. Amendola and D. T. Valentini,
Phys. Rev. D 66 (2002)  043528.


\bibitem{Zim} W. Zimdahl and D. Pavon, Phys. Lett. B 521 (2001) 133;\\ W. Zimdahl and D. Pavon, Gen. Rel. Grav. 35
(2003) 413;\\ L. P. Chimento, A. S. Jakubi, D. Pavon and W.
Zimdahl, Phys. Rev. D 67 (2003)  083513.

\bibitem{Seta1}
M. R. Setare, Eur. Phys. J. C 50 (2007) 991;\\ M. R. Setare, JCAP
0701 (2007) 023;\\ M. R. Setare, Phys. Lett. B 654 (2007) 1;\\
M. R. Setare, Phys. Lett. B 642  (2006) 421.

\bibitem{wang1} B. Wang, Y. Gong and E. Abdalla, Phys. Lett. B 624
(2005) 141;\\ B. Wang, C. Y. Lin. D. Pavon and E. Abdalla, Phys.
Lett. B 662 (2008) 1.

\bibitem{Oli} G. Olivares, F. Atrio, D. Pavon, Phys. Rev. D 71 (2005) 063523.

\bibitem{wang2} B. Wang, C. Y. Lin and E. Abdalla, Phys. Lett. B 637
(2005) 357;\\ M. R. Setare, Eur. Phys. J. C 52 (2007) 689;\\ Bin
Wang, et al., astro-ph/0607126.




\bibitem{spe} C. L. Bennett, et al.,  Astrophys. J. Suppl.
148 (2003) 1;\\ D. N. Spergel, Astrophys. J. Suppl. 148 (2003) 175;\\
M. Tegmark, et al., Phys. Rev. D 69 (2004) 103501;\\ U. Seljak, A.
Slosar, P. McDonald, JCAP 0610 (2006) 014;\\ D. N. Spergel, et
al., Astrophys. J. Suppl. 170 (2007) 377.


\bibitem{Sie} J. L. Sievers, et al., Astrophys. J. 591 (2003) 599;\\ C.B.
Netterfield, et al., Astrophys. J. 571 (2002) 604;\\ A. Benoit, et
al., Astron. Astrophys. 399 (2003) L25;\\ A. Benoit, et al.,
Astron. Astrophys. 399 (2003) L19.


\bibitem{Uzan} J. P. Uzan, U. Kirchner, G.F.R. Ellis, Mon. Not. R. Astron.
Soc. 344 (2003) L65;\\ A. Linde, JCAP 0305 (2003) 002;\\ M.
Tegmark, A. de Oliveira-Costa, A. Hamilton, Phys. Rev. D 68 (2003)
123523;\\ G. Efstathiou, Mon. Not. R. Astron. Soc. 343 (2003)
L95;\\ J. P. Luminet, J. Weeks, A. Riazuelo, R. Lehou, J. Uzan,
Nature 425 (2003) 593; \\ G. F. R. Ellis, R. Maartens, Class.
Quantum Grav. 21 (2004) 223.


\bibitem{Caldwell} R. R. Caldwell, M. Kamionkowski, astro-ph/0403003;\\ B. Wang,
Y. G. Gong, R. K. Su, Phys. Lett. B 605 (2005) 9.

\bibitem{Tsu} S. Tsujikawa, M. Sami, Phys. Lett. B 603 (2004) 113.

\bibitem{Dor} M. Doran, J. Jackel, Phys. Rev. D 66 (2002) 043519.

\end{thebibliography}
\end{document}